\documentclass[11pt]{article}
\usepackage{amsmath,enumerate,amsfonts,color,url,amssymb,ntheorem,cite}

{\catcode `\@=11 \global\let\AddToReset=\@addtoreset}
\AddToReset{equation}{section}

\AddToReset{figure}{section}

\newcounter{mnotecount}[section]
\renewcommand{\themnotecount}{\thesection.\arabic{mnotecount}}

\DeclareFontFamily{OT1}{rsfs}{}
\DeclareFontShape{OT1}{rsfs}{m}{n}{ <-7> rsfs5 <7-10> rsfs7 <10-> rsfs10}{}
\DeclareMathAlphabet{\mycal}{OT1}{rsfs}{m}{n}

\definecolor{MDB}{rgb}{0,0.08,0.45}
\definecolor{MyDarkBlue}{rgb}{0,0.08,0.45}

\definecolor{MLM}{cmyk}{0.1,0.8,0,0.1}
\definecolor{MyLightMagenta}{cmyk}{0.1,0.8,0,0.1}

\definecolor{HP}{rgb}{1,0.09,0.58}

\newcommand{\jlcax}[1]{}
\newcommand{\eean}{\nonumber\end{eqnarray}}

\newcommand{\Uone}{\mathrm{U(1) }}%

\newcommand{\kk}[1]{}%{\mnote{{\bf If we consider the KK case:} #1}}

\newcommand{\beq}{\begin{equation}}

\newcommand{\FS}       %{F_1} %
                  {F}
\newcommand{\HS} %{F_2}
       {H_{\mbox{\scriptsize volume}}}

{\ptc{this should be removed in the oberwolfach version}}%

\newcommand{\mcA}{{\mycal A}}%
\newcommand{\eeal}[1]{\label{#1}\end{eqnarray}}
\newcommand{\bed}{\begin{deqarr}}
\newcommand{\eed}{\end{deqarr}}
\newcommand{\bedl}[1]{\begin{deqarr}\label{#1}}
\newcommand{\eedl}[2]{\arrlabel{#1}\label{#2}\end{deqarr}}

\newcommand{\bel}[1]{\begin{equation}\label{#1}}
\newcommand{\bea}{\begin{eqnarray}}
\newcommand{\bean}{\begin{eqnarray}\nonumber}
\newcommand{\beal}[1]{\begin{eqnarray}\label{#1}}
\newcommand{\eea}{\end{eqnarray}}

\newcommand{\be}{\begin{equation}}
\newcommand{\eeq}{\end{equation}}
\newcommand{\ee}{\end{equation}}
\newcommand{\beqa}{\begin{eqnarray}}
\newcommand{\eeqa}{\end{eqnarray}}
\newcommand{\beqan}{\begin{eqnarray*}}
\newcommand{\eeqan}{\end{eqnarray*}}
\newcommand{\ba}{\begin{array}}
\newcommand{\ea}{\end{array}}

\newcommand{\scri}{{\mycal I}}%
\newcommand{\scrip}{\scri^{+}}%

\newcommand{\mnote}[1]%{}
{\protect{\stepcounter{mnotecount}}$^{\mbox{\footnotesize
$%\!\!\!\!\!\!\,
\bullet$\themnotecount}}$ \marginpar{%\color{red}%
\raggedright\tiny\em
$\!\!\!\!\!\!\,\bullet$\themnotecount: #1} }

\newcommand{\warn}[1]%{}%{}
{\protect{\stepcounter{mnotecount}}$^{\mbox{\footnotesize
$%\!\!\!\!\!\!\,
\bullet$\themnotecount}}$ \marginpar{%\color{red}%
\raggedright\tiny\em
$\!\!\!\!\!\!\,\bullet$\themnotecount: {\bf Warning:} #1} }

\newcommand{\R}{\mathbb R}

\newcommand{\ptc}[1]{\mnote{{\bf ptc:}#1}}

\newcommand{\beqar}{\begin{deqarr}}
\newcommand{\eeqar}{\end{deqarr}}

\newcommand{\beaa}{\begin{eqnarray*}}
\newcommand{\eeaa}{\end{eqnarray*}}

\def\RR{{\mathbb R}}

\def\CC{{\mathbb C}}

\def\FS{{\mathfrak S}}

\def\lMext{{M_{\rm ext}}}

\def\Msh{{M_\sharp}}
\def\Csh{{\CC_\sharp}}
\def\Omsh{{\Omega_\sharp}}
\def\Gsh{{G_\sharp}}
\def\Qsh{{Q_\sharp}}
\def\Ush{{U_\sharp}}
\def\ash{{\alpha_\sharp}}
\def\rsh{{\rho_\sharp}}
\def\zsh{{z_\sharp}}
\def\zetash{{\zeta_\sharp}}
\def\mcAsh{{\mcA_{\sharp}}}

\def\USSchw{{U_{S,{\rm\scriptsize Schw}}}}

\def\ringM{{\mathring{M}}}

\newtheorem{Theorem} {\sc  Theorem\rm} [section]

\newtheorem{Proposition} [Theorem] {\sc  Proposition\rm}

\newtheorem{corollary} [Theorem] {\sc  Corollary\rm}

\newtheorem{proposition} [Theorem] {\sc  Proposition\rm}
\newtheorem{theorem}[Theorem]{\sc  Theorem\rm}

\theorembodyfont{\upshape}

\newtheorem{remark}[Theorem]{\sc Remark\rm}

\def\bproof{\noindent{\sc Proof:\;}}
\def\eproof{\hfill$\square$\medskip}

\newcounter{marnote}

\newcommand\divop{\textrm{div}\,}

\begin{document}

\title{A lower bound for the mass of axisymmetric connected black hole data sets}
\author{Piotr T. Chru\'{s}ciel \thanks{Gravitational Physics, University of Vienna}~ and Luc Nguyen \thanks{Department of Mathematics, Princeton University}}
\maketitle

\begin{abstract}
We present a generalisation of the Brill-type proof of
positivity of mass for axisymmetric initial data  to initial
data sets with black hole boundaries. The argument leads to a strictly positive
lower bound for the mass of simply connected, connected
axisymmetric black hole data sets in terms of the mass of a
reference Schwarzschild metric.
\end{abstract}

\tableofcontents

%----------------------------------------------------------------------------%
\section{Introduction}

In \cite{ChUone}, the first author extended the validity of the
axisymmetric positive mass theorems  of Brill \cite{Brill59},
Moncrief (unpublished), Dain (unpublished) and Gibbons and
Holzegel \cite{GibbonsHolzegel} to all asymptotically flat
initial data on $\RR^3$ invariant under a $\Uone$--action with
positive Ricci scalar. The object of this work is to show how
to adapt the analysis to the case where black hole boundaries
are present in the initial data. This leads  to a
strictly positive lower bound for the mass for initial data
sets containing a connected ``non-degenerate
horizon''.

Let $(M,g)$ be a simply connected  three dimensional Riemannian
manifold with boundary $\partial M$ which admits a Killing
vector field with periodic orbits and is the union of a compact
set and of one asymptotically flat end.

By \cite{ChUone}, $g$ admits a global coordinate system in
which the metric takes the form
\begin{equation}
ds^2 = e^{-2\Ush + 2\ash}(d\rsh^2 + d\zsh^2) + \rsh^2\,e^{-2\Ush}(d\varphi + \rsh\,\bar B\,d\rsh + \bar A\,d\zsh)^2
	\;,
\label{metric}
\end{equation}
where $\partial_\varphi$ is the Killing vector field,
$\varphi\in[0,2\pi)$, the coordinates $(\rsh,\zsh)$ cover
$([0,\infty) \times \RR) \setminus \mathring{K}$ for some
compact set $K$ whose intersection with the axis $\{\rsh = 0\}$
is connected and non-empty. Here $\mathring K$ denotes the
interior of $K$. The choice of $K$ is never unique.
However, we show in the present paper that it cannot be
arbitrary either. In fact, if one lets $K'$ be the compact set
obtained by adjoining to $K$ its reflection about the $\rsh$
axis in the $(\rsh,\zsh)$-plane, then the logarithmic capacity
of $K'$ (with respect to the $(\rsh,\zsh)$-plane) depends
uniquely on the geometry of $(M,g)$.

The above property of $K$ leads to two canonical choices of
$K$: one can use either a line segment of length $2m_1$ on the
$\rsh$-axis, or a half-disc of radius $\frac{m_1}{2}$ centered
on the axis, where $m_1$ is twice the logarithmic capacity of
$K'$. In the stationary vacuum case, those coordinate systems
are respectively known as Weyl coordinates and isotropic (or
spherical) coordinates. In the static case, i.e.\ $(M,g)$ is a
Schwarzschild slice, $m_1$ coincides with the Schwarzschild
mass. For the maximal slice of the Kerr metric, $m_1 =
\sqrt{m^2 - a^2}$.

The main result of the present paper is as follows.
\begin{theorem}\label{MainThm}
Let $(M,g)$ be a smooth simply connected three-dimensional
manifold which has a smooth connected compact boundary $\partial M$, is asymptotically flat with one end and
satisfies \eqref{FO-1} for some $k \geq 5$ and \eqref{FO-2}. Furthermore, assume that $(M,g)$ admits a Killing vector field with periodic orbits. If $M$ has non-negative scalar curvature and if the  mean curvature of $\partial M$  with respect to the normal pointing towards $M$ is non-positive, then the ADM mass of
$(M,g)$ satisfies
\begin{equation}
m > \frac{\pi}{4}\,m_1
	\;,
\label{MassBound}
\end{equation}
where $m_1$ is the positive constant obtained in Theorems \ref{RSRepThm} and \ref{RepThm}.
\end{theorem}

Even though the constant $m_1$ is uniquely determined, it should be admitted that it is not easy to
determine $m_1$ if the metric is not given directly in the
coordinate system \eqref{MetricS::CanF} or \eqref{Metric::CanF}. In the general case, one needs to solve a PDE on $M$, and then $m_1$ can be read off from the asymptotic behaviour of the solution at infinity, see Proposition \ref{m1Value?}.

We note that the simple-connectedness of $M$ would be a consequence of the topological
censorship theorem of~\cite{Galloway:fitopology} if $M$ were a
Cauchy hypersurface for $J^+(M)\cap J^-(\scrip)$. We are grateful to
G.~Galloway for pointing this out.

A satisfactory generalisation of our result to degenerate horizons would require a thorough understanding of the behaviour of the metric near such horizons, a problem which is widely unexplored. We simply note that positivity of $m$ is easily established by similar methods if one assumes, e.g., that   $(M,g)$ has no boundary but contains instead suitably defined asymptotically cylindrical ends.
In this case, whenever a twist potential exists one further has the stronger Dain-type inequality controlling the mass from below in terms of a  positive quantity, which equals the square root of the length of the  angular-momentum for connected configurations.

We conjecture that the sharp inequality is
\begin{equation}
m \ge m_1
	\;,
\label{MassBoundSharp}
\end{equation}
with equality if and only if $M$ is a time-symmetric Cauchy hypersurface for the d.o.c. of the Schwarzschild-Kruskal-Szekeres space-time.

Ideally, one would like to obtain a simple proof of the Penrose
inequality in the current setting, but we have not been able to
achieve this. We make some comments about that in the appendix.

We note the recent paper~\cite{ADG}, where a lower bound for
minimal-surface area in terms of  angular-momentum  is
established under a set of restrictive conditions; see
also~\cite{HCAUniversalInequality}. Our construction of global coordinates is relevant for the
analysis in~\cite{ADG}.

%----------------------------------------------------------------------------%
\section{Axisymmetric black hole data sets}

Let $(M,g)$ be a three-dimensional smooth simply connected
manifold with a smooth connected compact boundary $\partial M$.
On $(M,g)$, we assume that there is a Killing vector field
$\eta$ with periodic orbits, among which the principal ones are
assumed, without loss of generality, to have period $2\pi$.

$(M,g)$ will be assumed to have {\it one} asymptotically flat end in the
usual sense that there exists a region $\lMext \subset M$
diffeomorphic to $\RR^3\setminus B_R$, where $B_R$ is a
coordinate ball of radius $R$, such that in local coordinates
on $\lMext$ obtained from $\RR^3\setminus B_R$ the metric
satisfies the fall-off conditions, for some $k \geq 1$,
\begin{align}
&g_{ij} - \delta_{ij} = o_k(r^{-1/2})
	\;,\label{FO-1}\\
&\partial_k g_{ij} \in L^2(\lMext)
	\;,\label{FO-2}\\
&R^i{}_{jkl} = o(r^{-5/2})
	\;,\label{FO-3}
\end{align}
where we write $f = o_k(r^{\mu})$ if $f$ satisfies
\bel{oldef}
\partial_{k_1 \ldots k_l} f = o(r^{\mu - l}) \text{ for } 0 \leq l \leq k
	\;.
\ee

The boundary behaviour near the event horizon of all functions
of interest has been established in~\cite{ChCo} in a closely
related context. For completeness, and to make it clear that it
applies to our setting, we outline the analysis
of~\cite{ChCo} in what follows.

%++++++++++++++++++++++++++++++++++++++++%
\subsection{Reduction to the boundaryless case}

Since $M$ is simply connected and $\partial M$ is connected,
\cite[Lemma 4.9]{Hempel} implies that
\begin{equation}
\text{$\partial M$  has the topology of a $2$-sphere,}
\label{SphericalBdry}
\end{equation}
and can be filled by a ball, say $B_{\sharp}$. Moreover, the
metric $g$ and the $U(1)$-action induced by $\eta$ extend to
the extended manifold $\Msh := M \cup B_\sharp$.

Let
\[
\mcA := \{p \in M: g(\eta,\eta)|_p = 0\}
	\;, \qquad \mcAsh := \{p \in \Msh: g(\eta,\eta) |_p= 0\}
	\;.
\]
Denote by $\Qsh := \Msh/U(1)$, $Q := M/U(1)$ and $\partial
M/U(1)$ the collections of the orbits of the group of
isometries generated by $\eta$ on $\Msh$, $M$ and $\partial M$,
respectively. It is known  that  $\Qsh$ is a manifold with
boundary $\mcAsh/U(1) \approx  \mcAsh$, while $Q$ is a manifold
with boundary
$$
 (\mcA/U(1)) \cup (\partial M/U(1)) \approx  \mcA
 \cup (\partial M/U(1))
  \;.
$$
As $\partial M$ is a sphere and
invariant under $U(1)$, it contains exactly two fixed points,
say $p_n$ and $p_s$, of $U(1)$. In other words,
\[
\partial M \cap \mcAsh = \{p_n, p_s\}
	\;.
\]
It is readily seen that $\partial M/U(1)$ is a smooth curve in
$\Qsh$ with endpoints $p_n$ and $p_s$. Additionally, at
both $p_n$ and $p_s$, $\partial M/U(1)$ intersects $\mcAsh$ at
a right angle.

In the sequel, we will assume that \eqref{FO-1} holds for some
$k \geq 5$. Note that this implies \eqref{FO-3}. By
\cite[Theorem 2.7]{ChUone}, the metric $g$ on $\Msh$ admits the
following representation
\begin{equation}
g = e^{-2\Ush + 2\ash}(d\rsh^2 + d\zsh^2) + \rsh^2\,e^{-2\Ush}(d\varphi + \rsh\,\bar B\,d\rsh + \bar A\,d\zsh)^2
	\;,\label{MetricSh::CanF}
\end{equation}
where $\partial_\varphi$ is the rotational Killing vector
field, and $\Msh$ can be identified with $\RR^3$ on which
$(\rsh,\zsh,\varphi)$ are its cylindrical coordinates.
Furthermore, $\Ush$, $\ash$, $\bar B$ and $\bar A$ are smooth
functions on $\Msh$ which are $\varphi$-independent and satisfy
$\ash = 0$ whenever $\rsh = 0$ and
\begin{multline}
\Ush = o_{k-3}(r_\sharp^{-1/2})
	\;,	\qquad	\ash = o_{k-4}(r_\sharp^{-1/2})
	\;,\\
		\qquad	\bar B = o_{k-3}(r_\sharp^{-5/2})
	\;,	\qquad	\bar A = o_{k-3}(r_\sharp^{-3/2}) \text{ for } r_\sharp = \sqrt{\rsh^2 + \zsh^2} \rightarrow \infty
	\;.\label{FOMsh::CanF}
\end{multline}

It is useful to consider the manifolds $\Qsh^{(2)}$ and
$Q^{(2)}$ obtained by doubling $\Qsh$ and $Q$ along
$\mcAsh/U(1)$ and $\mcA/U(1)$, respectively. Naturally, $Q^{(2)}$
injects into $\Qsh^{(2)}$. In addition, $\rsh$ and $\zsh$ can
be extended naturally to $\Qsh^{(2)}$ so as to make $\rsh$ an
odd function about $\mcAsh$ while $\zsh$ an even function. Then
$\Qsh^{(2)}$ can be identified with the complex plane $\Csh :=
\{\zetash := \rsh + i\zsh\}$ in which
$$\Qsh \approx \{\zetash: {\rm
Re}\, \zetash \geq 0\}
 \;,
 \qquad \mcAsh  \approx \{\zetash: {\rm Re}\,\zetash
= 0\}
 \;,
$$
where $\approx$ denotes ``diffeomorphic to". This implies in particular that $\Qsh$ has a natural complex structure.
Furthermore, in this picture, $Q^{(2)}$ is an unbounded (open)
subset, denoted by $\Omsh$, of $\Csh$ whose boundary is a
smooth connected closed curved,
$$Q \approx \Omsh \cap \{\zetash:
{\rm Re}\,\zetash \geq 0\} \ \mbox{ and } \
 \mcA \approx \Omsh \cap \{\zetash:
 {\rm Re}\,\zetash = 0\}
 \;.
$$

%++++++++++++++++++++++++++++++++++++++++%
\subsection{Pseudo-spherical coordinates}

We proceed to modify $(\rsh,\zsh,\varphi)$ to a coordinate
system $(\rho_S,z_S,\varphi)$ on $M$ such that $\partial M$
corresponds to a sphere $\{\rho_S^2 + z_S^2 = {\rm const}\}$.
An approach to achieve this is to follow the procedure in
\cite{ChCo} to first construct Weyl coordinate functions and
then transform them to the desired form. We present here a
simpler approach, directly tied to the theory of conformal
mappings. As will be seen, this also provides an alternative to
the construction of Weyl coordinates in \cite{ChCo}.

Without loss of generality, we assume that
$\Omsh$ does not contain the origin of $\Csh$. Let $\Theta$
denote the inversion map of $\Csh$ about the unit circle
$\partial D_\sharp(0,1)$ and define $\Gsh = \Theta(\Omsh) \cup
\{0\}$. Note that as $\partial\Omsh$ is a smooth simple closed
curved, so is $\partial\Gsh$. This implies that $\Gsh$ is
simply connected. Let $h_1$ be the solution to the problem
\[
\left\{\begin{array}{ll}
\Delta^\sharp h_1 = 0 & \text{ in } \Gsh
	\;,\\
h_1 = -\frac{1}{2}\log(\rsh^2 + \zsh^2) &\text{ on } \partial \Gsh
	\;,
\end{array}\right.
\]
where $\Delta^\sharp$ is the Laplace operator of $d\rsh^2 + d\zsh^2$, and $h_2$ be a harmonic conjugate of $h_1$, i.e. $h_2$ satisfies
\[
\partial_{\rsh} h_1 = \partial_{\zsh} h_2
	\;,\qquad \partial_{\zsh} h_2 = -\partial_{\rsh} h_1
	\;.
\]
Define
$$
 \Psi \equiv \psi_1 + i\psi_2 := \zetash\,\exp(h_1 + i\,h_2)
 \;;
$$
%.
recall that $\zetash = \rsh + i \zsh$. Evidently, $\Psi$ is
holomorphic, fixes the origin, and maps $\partial\Gsh$  to the
unit circle $\partial D_\sharp(0,1)$. Furthermore, by the definitions of $h_1$, $h_2$ and standard
elliptic theory, $\Psi \in C^\infty(\bar\Gsh)$.

We claim that $\Psi$ is ``the'' Riemann map which maps $\Gsh$
one-to-one and onto the unit disc $D_\sharp(0,1)$. Indeed, let
$\tilde \Psi$ be a Riemann  map of $\Gsh$ which fixes the
origin. Then $\tilde \Psi = \zetash \,\tilde H$ for some
holomorphic function $\tilde H$. Additionally, as $\tilde\Psi$
is one-to-one, $\tilde H$ is nowhere vanishing. Since $\Gsh$ is
simply connected, this implies $\tilde H = \exp\,\tilde h$ for
some holomorphic function $\tilde h$. As $\tilde\Psi(\partial
\Gsh) \subset \partial D_\sharp(0,1)$, it follows that
\[
{\rm Re}\,\tilde h = -\log|\zetash| = -\frac{1}{2}\log(\rsh^2 + \zsh^2) \text{ on } \partial \Gsh
	\;.
\]
By  uniqueness of solutions of the Laplace equation, we thus
have ${\rm Re}\,\tilde h \equiv h_1$, which implies ${\rm
Im}\,\tilde h \equiv h_2 + C$ for some constant $C$. The claim
follows.

As a Riemann map, $\Psi$ has an inverse $\Psi^{-1}:
D_\sharp(0,1) \rightarrow  \Gsh$. Since $\Gsh$ is a Jordan
domain, $\Psi^{-1}$ extends to a homeomorphism of the closed
domains thanks to Carath\'eodory theorem (see e.g.
\cite[Theorem~14.19]{Rudin}).  We claim that this extension is
of $C^\infty(\bar D_\sharp(0,1))$-differentiability class, and
in fact is a diffeomorphism up-to-boundary. By the Inverse
Function Theorem, it suffices to show that $\Psi'$ is nowhere
vanishing in $\bar\Gsh$. Furthermore, since $\Psi$ is
holomorphic and one-to-one in $\Gsh$, it suffices to show that
$\Psi'$ does not vanish on $\partial \Gsh$. Consider a point $p
\in \partial \Gsh$ and let $q = \Psi(p) \in \partial
D_\sharp(0,1)$. Without loss of generality, we can assume that
$q = -i$. Pick a $\delta
> 0$ sufficiently small such that
\begin{equation}
\Psi(\Gsh \cap D_\sharp(p,\delta)) \subset \bar D_\sharp(0,1) \cap D_\sharp(q, \frac{1}{10})
	\;.\label{RM::Nbhd}
\end{equation}
Since $\psi_1^2 + \psi_2^2=1$ on $\partial\Gsh$ we find that,
near $p$, the function
\[
F := \psi_2 + \sqrt{1 - \psi_1^2}
%	\;.
\]
satisfies
\begin{align*}
&F > 0 \text{ in } \Gsh \cap D_\sharp(p,\delta)
	\;,\\
&F = 0 \text{ on } \partial \Gsh \cap D_\sharp(p,\delta)
    \;.
\end{align*}
Since $\psi_1$ and $\psi_2$ are harmonic, we have
\[
\Delta^\sharp F = -\frac{1}{(1 - \psi_1^2)^{3/2}}\,|\nabla\psi_1|^2 \leq 0 \text{ in }  \Gsh \cap D_\sharp(p,\delta)
	\;.
\]
It hence follows from the Hopf lemma that
\[
\partial_\nu \psi_2(p) = \partial_{\nu} F(p) > 0
	\;,
\]
where $\partial_\nu$ is the derivative in the direction of the
inward pointing normal, which gives $|\Psi'(p)| \geq
|\partial_\nu \psi_2(p)
> 0$. Since $p$ is arbitrary, we thus conclude that $\Psi'$ is
always non-zero on $\partial \Gsh$ and so on $\bar\Gsh$, whence
the claim.

\medskip

Note that we have recovered the Kellogg-Warschawski theorem (see \cite[Theorem 3.6]{PommerenkeBook} or the original papers \cite{Kellogg,Warschawski1,Warschawski2} of Kellogg and of Warschawski):

\begin{proposition}
 \label{PLuc} Let $G \subset \CC$ be a simply connected bounded domain whose boundary $\partial G$ is $C^{k,\alpha}$-regular for some $k \geq 2$, $0 < \alpha < 1$, and $\Psi: G \rightarrow D(0,1)$ its Riemann map. Then $\Psi$ extends to a map in $C^{k,\alpha}(\bar G)$ and $\Psi^{-1}$ extends to a map in $C^{k,\alpha}(\bar D(0,1))$.
\end{proposition}

Define
\begin{equation}
\rho_S + i\,z_S = \zeta_S := \Theta^{-1} \circ \Big(\frac{1}{|\Psi'(0)|}\Psi\Big) \circ \Theta
	\;.\label{zetaSCmp}
\end{equation}
Then $(\rho_S,z_S)$ maps $\Omsh$ one-to-one and onto $\CC_S \setminus \bar D(0,\frac{m_1}{2})$, where  we use the
symbol $\CC_S$ to denote the complex plane coordinatized by
$(\rho_S, z_S)$, and where
$$
 m_1 = 2|\Psi'(0)|
$$
is twice the logarithmic capacity of $\partial\Omsh$. The
constant $m_1$ is related to the Robin constant
$\gamma(\partial\Omsh)$ of the boundary of $\partial \Omsh$ by
\begin{equation}
m_1 = 2\,\exp(- \gamma(\partial\Omsh))
	\;.\label{CF::m1Def}
\end{equation}
Note also that by construction, as $\rsh^2 +
\zsh^2 \rightarrow \infty$, there holds
\begin{equation}
(\rho_S,z_S) - (\rsh,\zsh) = O_l((\rsh^2 + \zsh^2)^{-1/2})	\;, \qquad l \geq 0
	\;,\label{CoordSAsym}
\end{equation}
where $O_l$ is defined in a way analogous to \eqref{oldef}.

We show that $m_1$ is uniquely determined by $(M,g)$, i.e.
independent of how we form $\Msh$. To see this, let $\tilde
\Msh = M \cup \tilde B_\sharp$ be a different way of extending
$M$. In $\Msh$ and $\tilde \Msh$, the regions representing $M$
are isometric. Hence, if
$(\tilde\rsh,\tilde\zsh,\tilde\varphi)$ is the counterpart in
$\tilde \Msh$ of $(\rsh,\zsh,\varphi)$, then, by
\eqref{MetricSh::CanF}, the map $T: (\rsh,\zsh) \mapsto
(\tilde\rsh,\tilde\zsh)$ gives a conformal transformation of
$Q^{(2)}$. Furthermore, by \eqref{FOMsh::CanF}, we also have
\bel{24V10.1}
 \left|\frac{\partial(\tilde\rsh,\tilde\zsh)}{\partial(\rsh,\zsh)}\right|
 = 1 + O((\rsh^2 + \zsh^2)^{-1/2}) \text{ as } \rsh^2 + \zsh^2 \rightarrow \infty
	\;. \ee
Hence, if $Q^{(2)}$ is represented by $\tilde\Omega_\sharp$ in
the complex plane $\tilde\CC_\sharp$ parameterized by
$(\tilde\rsh,\tilde\zsh)$, then $T$ defines naturally a
bijection of $\Omsh$ and $\tilde\Omega_\sharp$, with $T(\infty)
= \infty$ and (by \eqref{24V10.1}) $|T'(\infty)| = 1$. It then
follows that $\partial \Omsh$ and $\partial
\tilde\Omega_\sharp$ have the same logarithmic capacity, and so
$m_1$ is independent of the way that the metric has been
extended to $B_\sharp$.

Next, by the
uniqueness property of the Laplace equation with Dirichlet
boundary data, $h_1$ is even in the $\rsh$-variable, which implies
that, after shifting by a constant, $h_2$ is odd in the
$\rsh$-variable. Using this, one can check that $\rho_S$ is odd
while $z_S$ is even in the $\rsh$-variable. In particular,
$\rho_S$ vanishes on $\{\rsh = 0\} \cap \Omsh \approx
\mcA/U(1)$. This implies that $\rho_S^2$ is a smooth function
which vanishes on $\mcA$ and is even in the $\rsh$-variable.
Thus, there is a smooth function $\chi$ of $(\rsh^2,\zsh)$ such
that
\begin{equation}
\rho_S^2 = \rsh^2\,\chi(\rsh^2,\zsh)
	\;.\label{RSFtrztn}
\end{equation}
Furthermore, as
$$
 (\partial_{\rsh}\rho_S)^2 + (\partial_{\zsh}\rho_S)^2
  =
  \left|\frac{\partial(\rho_S,z_S)}{\partial(\rsh,\zsh)}\right|
$$
is nowhere vanishing in $\overline{\Omsh}$ by Proposition
\ref{PLuc}, we also have
\begin{equation}
\chi(\rsh^2,\zsh) > 0 \text{ along } \{\rsh = 0\} \cap \overline{\Omsh}\;,
\text{ and so in } \overline{\Omsh}
	\;.\label{NontrivialFtrztn}
\end{equation}

We thus have:
\begin{theorem}
 \label{RSRepThm}
Let $(M,g)$ be a three-dimensional smooth simply connected manifold with a smooth connected compact boundary $\partial M$ and assume that $(M,g)$ admits a Killing vector field with periodic orbits. Furthermore, assume that $(M,g)$ has one asymptotically flat end where it satisfies \eqref{FO-1} for some $k \geq 5$. Then there exists a unique $m_1 > 0$ such that $M$ is diffeomorphic to $\RR^3\setminus B(0,\frac{m_1}{2})$, and, in cylindrical-type coordinates $(\rho_S,z_S,\varphi)$ on $\RR^3$, $g$ takes the form
\begin{equation}
g = e^{-2U_S + 2\alpha_S}(d\rho_S^2 + dz_S^2) + \rho_S^2\,e^{-2U_S}(d\varphi + \rho_S\,\bar B_S\,d\rho_S + \bar A_S\,dz_S)^2
	\;,\label{MetricS::CanF}
\end{equation}
where $\partial_\varphi$ is the rotational Killing vector
field, $U_S$, $\alpha_S$, $\bar B_S$ and $\bar A_S$ are smooth
functions on $M$ which are $\varphi$-independent and satisfy
$\alpha_S = 0$ whenever $\rho_S = 0$ and
\begin{multline}
U_S = o_{k-3}(r_S^{-1/2})
	\;,	\qquad	\alpha_S = o_{k-4}(r_S^{-1/2})
	\;,\\
		\qquad	\bar B_S = o_{k-3}(r_S^{-5/2})
	\;,	\qquad	\bar A_S = o_{k-3}(r_S^{-3/2}) \ \mbox{ \rm for }\ r_S = \sqrt{\rho_S^2 + z_S^2} \rightarrow \infty
	\;.\label{FOMS::CanF}
\end{multline}
\end{theorem}

%++++++++++++++++++++++++++++++++++++++++%
\subsection{Weyl coordinates}

We next construct the Weyl coordinates $(\rho,z,\varphi)$ so that $\rho$ vanishes on both the rotation axis $\mcA$ and the boundary $\partial M$. This can be done using a (rotated)
Joukovsky transformation,
\begin{equation}
\rho + i\,z = \zeta := \zeta_S - \frac{m_1^2}{4\zeta_S}
	\;.\label{zetaCmp}
\end{equation}
Componentwise, we have
\begin{equation}
\rho = \frac{\rho_S\,\big[\rho_S^2 + z_S^2 - \frac{m_1^2}{4}\big]}{\rho_S^2 + z_S^2}
	\;, \qquad z = \frac{z_S\,\big[\rho_S^2 + z_S^2 + \frac{m_1^2}{4}\big]}{\rho_S^2 + z_S^2}
	\;.\label{zetaCom}
\end{equation}
We now check that the map $\zeta_S \mapsto \zeta$ maps $\CC_S
\setminus \bar D(0,\frac{m_1}{2})$ one-to-one and onto
$\CC\setminus I$ where $I = \{i\,z: -m_1 \leq z \leq m_1\}$. In
view of \eqref{zetaCom}, to invert the map it suffices to solve
for $|\zeta_S|
> \frac{m_1}{2}$. First, note that by \eqref{zetaCmp}
\begin{equation}
\zeta \pm i\,m_1 = \frac{1}{\zeta_S}\big(\zeta_S \pm i\,\frac{m_1}{2}\big)^2
	\;.\label{zeS-ze-1}
\end{equation}
It follows that
\[
(\zeta + i\,m_1)(\bar\zeta + i\,m_1)
	= \frac{1}{|\zeta_S|^2}\Big(|\zeta_S|^2 - \frac{m_1^2}{4} + i\,\frac{m_1}{2}\,(\zeta_S + \bar\zeta_S)\Big)^2
	\;.
\]
Taking the real part and recalling \eqref{zetaCom} we get
\begin{align}
|\zeta|^2 - m_1^2
	&= \frac{1}{|\zeta_S|^2}\Big[\big(|\zeta_S|^2 -
 \frac{m_1^2}{4}\big)^2 - m_1^2\,\rho_S^2\Big]\nonumber\\
	&= \frac{\big(|\zeta_S|^2 - \frac{m_1^2}{4}\big)^2}{|\zeta_S|^2} - \frac{m_1^2\,|\zeta_S|^2}{\big(|\zeta_S|^2 - \frac{m_1^2}{4}\big)^2}\,\rho^2
	\;.\label{zeS-ze-2}
\end{align}
This implies that
\begin{equation}
\frac{\big(|\zeta_S|^2 - \frac{m_1^2}{4}\big)^2}{|\zeta_S|^2} = \frac{1}{2}\Big[|\zeta|^2 - m_1^2 + \sqrt{(|\zeta|^2 - m_1^2)^2 + 4\,m_1^2\,\rho^2}\Big]
	\;.\label{zeS-ze-3}
\end{equation}
As $|\zeta_S| > \frac{m_1}{2}$, we thus get
\begin{equation}
\frac{|\zeta_S|^2 - \frac{m_1^2}{4}}{|\zeta_S|} = \frac{1}{\sqrt{2}}\Big[|\zeta|^2 - m_1^2 + \sqrt{(|\zeta|^2 - m_1^2)^2 + 4\,m_1^2\,\rho^2}\Big]^{1/2}
	\;,\label{zeS-ze-x}
\end{equation}
which implies
\begin{multline}
|\zeta_S| = \frac{1}{2\sqrt{2}}\Big\{\Big[|\zeta|^2 - m_1^2 + \sqrt{(|\zeta|^2 - m_1^2)^2 + 4\,m_1^2\,\rho^2}\Big]^{1/2}\\
		+ \Big[|\zeta|^2 + m_1^2 + \sqrt{(|\zeta|^2 - m_1^2)^2 + 4\,m_1^2\,\rho^2}\Big]^{1/2}\Big\}
	\;.\label{zeS-ze-4}
\end{multline}
>From what has been said we see that the map $\zeta_S \mapsto
\zeta$ maps $\CC_S \setminus \bar D(0,\frac{m_1}{2})$
one-to-one and onto $\CC\setminus I$. In fact, in view of
\eqref{zetaCom}, \eqref{zeS-ze-3} and \eqref{zeS-ze-x}, its
inverse is given by
\begin{equation}
\rho_S = (\mu^{1/2} + 1)\frac{\rho}{2}
	\;,\qquad z_S = (\mu^{-1/2} + 1)\frac{z}{2}
	\;,\label{zeS-ze-Inverse}
\end{equation}
where
\begin{equation}
\mu = \frac{|\zeta|^2 + m_1^2 + \sqrt{(|\zeta|^2 - m_1^2)^2 + 4\,m_1^2\,\rho^2}}{|\zeta|^2 - m_1^2 + \sqrt{(|\zeta|^2 - m_1^2)^2 + 4\,m_1^2\,\rho^2}}
	\;.\label{zeS-ze-y}
\end{equation}

Recall that $\rho_S$ is odd in the $\rsh$-variable and so vanishes on $\mcA/U(1)$. Thus, by \eqref{zetaCom}, $\rho$ also vanish on $\mcA/U(1)$. Also by \eqref{zetaCom}, $\rho$ vanishes on $\partial M/U(1)$.
Moreover, by \eqref{zetaCmp}, as $\rho_S^2 +
z_S^2 \rightarrow \infty$, there holds
\begin{equation}
(\rho,z) - (\rho_S,z_S) = O_l((\rho_S^2 + z_S^2)^{-1/2})	\;, \qquad l \geq 0
	\;.\label{CoordAsym}
\end{equation}

It thus follows that:
\begin{theorem}
 \label{TLuc}
  The map
$$
 (\rsh,\zsh)\mapsto (\rho,z)
$$
defined by   \eqref{zetaSCmp} and \eqref{zetaCmp} provides a
holomorphic diffeomorphism from   $\mathring{Q}$ to the complex
half-plane $\{\zeta = \rho + i\,z: \rho > 0\}$.
\end{theorem}

In the $(\rho,z,\varphi)$ coordinate system the metric $g$ on
$M$ admits again a representation of the form
\begin{equation}
g = e^{-2U + 2\alpha}(d\rho^2 + dz^2) + \rho^2\,e^{-2U}(d\varphi + \rho\,B_\rho\,d\rho + A_z\,dz)^2
	\;.\label{Metric::CanF}
\end{equation}
In the rest of this section, we will use Theorem~\ref{TLuc} to
study the regularity properties of the functions involved.

First, using $g_{\varphi\varphi} = \rho^2\,e^{-2U(\rho,z)} = \rho_S^2\,e^{-2U_S(\rho_S,z_S)}$ and \eqref{zetaCom}, the function $U$ is given by
\begin{equation}
U(\rho,z) := U_S(\rho_S,z_S) - \log\frac{\rho_S}{\rho} = U_S(\rho_S,z_S) - \log\frac{\rho_S^2 + z_S^2}{\rho_S^2 + z_S^2 - \frac{m_1^2}{4}}
	\;.\label{CF::UDef}
\end{equation}
Recalling \eqref{zeS-ze-x}, the above relation can be rewritten as
\begin{equation}
U(\rho,z) = \tilde U_S(\rho_S,z_S) + \frac{1}{2}\log\Big[\rho^2 + z^2 - m_1^2 + \sqrt{(\rho^2 + z^2 - m_1^2)^2 + 4\,m_1^2\,\rho^2}\Big]
	\label{CF::UDefV}
\end{equation}
for some $\tilde U_S \in C^\infty(\CC_S \setminus D(0,\frac{m_1}{2}))$.

Next, by \eqref{zetaCmp},
\[
d\rho^2 + dz^2 = \frac{\big|\zeta_S^2 + \frac{m_1^2}{4}\big|^2}{|\zeta_S|^4}(d\rho_S^2 + dz_S^2)
	\;.
\]
Thus, using \eqref{CF::UDef} and
\begin{equation}
e^{-2U(\rho,z) + 2\alpha(\rho,z)}(d\rho^2 + dz^2) = e^{-2U_S(\rho_S,z_S)  + 2\alpha_S(\rho_S,z_S)}(d\rho_S^2 + dz_S^2) \text{ in } \bar Q^{(2)}
	\;,\label{QM::-U+al}
\end{equation}
we get
\begin{align}
\alpha(\rho,z)
	&= U(\rho,z) - U_S(\rho_S,z_S) + \alpha_S(\rho_S,z_S) + \log\frac{|\zeta_S|^2}{\big|\zeta_S^2 + \frac{m_1^2}{4}\big|}
	\nonumber\\
	&= \alpha_S(\rho_S,z_S) + \log \frac{|\zeta_S|^2 - \frac{m_1^2}{4}}{\big|\zeta_S^2 + \frac{m_1^2}{4}\big|}
	\;.\label{CF::AlDef}
\end{align}
Recalling \eqref{zeS-ze-1}, we can rewrite \eqref{CF::AlDef} as
\begin{equation}
\alpha(\rho,z)
	= \tilde\alpha_S(\rho_S,z_S) + \frac{1}{2}\log \frac{\rho^2 + z^2 - m_1^2 + \sqrt{(\rho^2 + z^2 - m_1^2)^2 + 4\,m_1^2\,\rho^2}}{2\sqrt{\rho^2 + (z - m_1)^2}\sqrt{\rho^2 + (z + m_1)^2}}
	\label{CF::AlDefV}
\end{equation}
for some $\tilde\alpha_S \in C^\infty(\CC_S \setminus D(0,\frac{m_1}{2}))$. Also, as $\alpha_S$ vanishes on the axis $\mcA$, \eqref{CF::AlDef} implies that
\begin{equation}
\alpha(\rho,z) = 0 \text{ for } \rho = 0, z \notin [-m_1,m_1]
	\;.\label{CF::AxisReg}
\end{equation}

We also need to understand the behaviour of the metric functions $B_\rho$ and $A_z$,
keeping in mind that $\bar B_S$ and $ \bar A_S$ are smooth up-to-boundary.
Since
\[
\rho_S\,\bar B_S(\rho_S,z_S)\,d\rho_S + \bar A_S(\rho_S,z_S)\,dz_S = \rho\,B_\rho(\rho,z)\,d\rho + A_z(\rho,z)\,dz
	\;,
\]
$B_\rho$ and $A_z$ satisfy
\begin{align}
B_\rho(\rho,z)
	&= \frac 14
 \big[(\mu^{1/2} + 1)^2 + \frac{1}{2}(\mu^{-1/2} + 1)\,\mu_{,\rho}\,\rho\big]\bar B_S(\rho_S,z_S)\nonumber\\
		&\qquad\qquad - \frac{1}{4}\,\rho^{-1}\,\mu^{-3/2}\,\mu_{,\rho}\,z\,\bar A_S(\rho_S,z_S)
	\;,\label{CF::BrDef}\\
A_z(\rho,z)
	&= \frac{1}{8}(\mu^{-1/2} + 1)\,\mu_z\,\rho^2\,\bar B_S(\rho_S,z_S)\nonumber\\
		&\qquad\qquad + \frac 12
 \big[(\mu^{-1/2} + 1) - \frac{1}{2}\mu^{-3/2}\,\mu_{,z}\,z\big]\bar A_S(\rho_S,z_S)
	\;.\label{CF::AzDef}
\end{align}

We compute from \eqref{zeS-ze-y}:
% \eject
%
\begin{align*}
\mu_{,\rho}
	&= -\frac{4m_1^2\rho}{\sqrt{(|\zeta|^2 - m_1^2)^2 + 4\,m_1^2\,\rho^2}}\,
		\frac{|\zeta|^2 + m_1^2 + \sqrt{(|\zeta|^2 - m_1^2)^2 + 4\,m_1^2\,\rho^2}}{\big[|\zeta|^2 - m_1^2 + \sqrt{(|\zeta|^2 - m_1^2)^2 + 4\,m_1^2\,\rho^2}\big]^2}
	\;,\\
\mu_{,z}
	&= -\frac{4m_1^2z}{\sqrt{(|\zeta|^2 - m_1^2)^2 + 4\,m_1^2\,\rho^2}}\,
		\frac{1}{|\zeta|^2 - m_1^2 + \sqrt{(|\zeta|^2 - m_1^2)^2 + 4\,m_1^2\,\rho^2}}
	\;.
\end{align*}
Also, note that by \eqref{zeS-ze-2} and \eqref{zeS-ze-3},
\begin{align*}
|\zeta|^2 - m_1^2 + \sqrt{(|\zeta|^2 - m_1^2)^2 + 4\,m_1^2\,\rho^2}
	&= \frac{2\big(|\zeta_S|^2 - \frac{m_1^2}{4}\big)^2}{|\zeta_S|^2}
	\;,\\
|\zeta|^2 + m_1^2 + \sqrt{(|\zeta|^2 - m_1^2)^2 + 4\,m_1^2\,\rho^2}
	&= \frac{2\big(|\zeta_S|^2 + \frac{m_1^2}{4}\big)^2}{|\zeta_S|^2}
	\;,\\
\sqrt{(|\zeta|^2 - m_1^2)^2 + 4\,m_1^2\,\rho^2}
	&= \frac{\big(|\zeta_S|^2 - \frac{m_1^2}{4}\big)^2}{|\zeta_S|^2} + \frac{m_1^2\,|\zeta_S|^2}{\big(|\zeta_S|^2 - \frac{m_1^2}{4}\big)^2}\,\rho^2
	\;.
\end{align*}
Thus
\begin{align}
\mu_{,\rho}
	&= -\frac{2m_1^2\,|\zeta_S|^4\,\rho}{\big(|\zeta_S|^2 - \frac{m_1^2}{4}\big)^4 + m_1^2\,|\zeta_S|^4\,\rho^2}\,
		\frac{\big(|\zeta_S|^2 + \frac{m_1^2}{4}\big)^2}{\big(|\zeta_S|^2 - \frac{m_1^2}{4}\big)^2}
	\;,\label{CF::mu,rho}\\
\mu_{,z}
	&= -\frac{2m_1^2\,|\zeta_S|^4\,z}{\big(|\zeta_S|^2 - \frac{m_1^2}{4}\big)^4 + m_1^2\,|\zeta_S|^4\,\rho^2}
	\;.\label{CF::mu,z}
\end{align}
We also have, from \eqref{zeS-ze-y} and the before-last displayed equations, that
\begin{equation}
\mu
	= \left(\frac{|\zeta_S|^2 + \frac{m_1^2}{4}}{|\zeta_S|^2 - \frac{m_1^2}{4}}\right)^2
	\;.\label{CF::mu}
\end{equation}
Substituting \eqref{CF::mu,rho}, \eqref{CF::mu,z},
\eqref{CF::mu} and \eqref{zeS-ze-Inverse} into \eqref{CF::BrDef} and \eqref{CF::AzDef} we obtain
\begin{align*}
&B_\rho(\rho,z)\\
	&\qquad= \left[\frac{|\zeta_S|^4}{\big(|\zeta_S|^2 - \frac{m_1^2}{4}\big)^2}
		- \frac{m_1^2\,|\zeta_S|^2\,\rho_S^2\,\big(|\zeta_S|^2 + \frac{m_1^2}{4}\big)}{2\big(|\zeta_S|^2 - \frac{m_1^2}{4}\big)^2\big[\big(|\zeta_S|^2 - \frac{m_1^2}{4}\big)^2 + m_1^2\,\rho_S^2\big]}
			\right]\bar B_S(\rho_S,z_S)\nonumber\\
		&\qquad\qquad + \frac{m_1^2\,|\zeta_S|^2\,z_S}{2\big(|\zeta_S|^2 - \frac{m_1^2}{4}\big)\big[\big(|\zeta_S|^2 - \frac{m_1^2}{4}\big)^2 + m_1^2\,\rho_S^2\big]}\, \bar A_S(\rho_S,z_S)
	\;,\\
&A_z(\rho,z)\\
	&\qquad= -\frac{m_1^2\,\rho_S^2\,z_S}{2\big[\big(|\zeta_S|^2 - \frac{m_1^2}{4}\big)^2 + m_1^2\,\rho_S^2\big]}\,\bar B_S(\rho_S,z_S)\nonumber\\
		&\qquad\qquad + \left[\frac{|\zeta_S|^2}{|\zeta_S|^2 + \frac{m_1^2}{4}}
			- \frac{m_1^2\,z_S^2\,\big(|\zeta_S|^2 - \frac{m_1^2}{4}\big)}{2\big(|\zeta_S|^2 + \frac{m_1^2}{4}\big)\big[\big(|\zeta_S|^2 - \frac{m_1^2}{4}\big)^2 + m_1^2\,\rho_S^2\big]}\right]\bar A_S(\rho_S,z_S)
	\;.
\end{align*}
We thus write
\begin{align}
B_\rho(\rho,z)
	&= \frac{|\zeta_S|^4}{2\big(|\zeta_S|^2 - \frac{m_1^2}{4}\big)^2\big[\big(|\zeta_S|^2 - \frac{m_1^2}{4}\big)^2 + m_1^2\,\rho_S^2\big]}\,\tilde B_S(\rho_S,z_S)\nonumber\\
	&= \frac{1}{\rho^2 + z^2 - m_1^2 + \sqrt{(\rho^2 + z^2 - m_1^2)^2 + 4m_1^2\,\rho^2}} \times \nonumber\\
		&\qquad\qquad \times \frac{1}{\sqrt{(\rho^2 + z^2 - m_1^2)^2 + 4m_1^2\,\rho^2}}\,\tilde B_S(\rho_S,z_S)
	\;,\label{CF::BrDefV}\\
A_z(\rho,z)
	&= \frac{|\zeta_S|^2}{\big(|\zeta_S|^2 - \frac{m_1^2}{4}\big)^2 + m_1^2\,\rho_S^2}\,\tilde A_S(\rho_S,z_S)\nonumber\\
	&= \frac{1}{\sqrt{(\rho^2 + z^2 - m_1^2)^2 + 4m_1^2\,\rho^2}}\,\tilde A_S(\rho_S,z_S)
	\;,\label{CF::AzDefV}
\end{align}
where $\tilde B_S, \tilde A_S \in C^\infty(\CC_S \setminus D(0,\frac{m_1}{2}))$.

Finally, by \eqref{FOMS::CanF} and \eqref{CoordAsym} and the above regularity justification, we have
\begin{multline}
U = o_{k-3}(r^{-1/2})
	\;,	\qquad	\alpha = o_{k-4}(r^{-1/2})
	\;,\\
		\qquad	B_\rho = o_{k-3}(r^{-5/2})
	\;,	\qquad	A_z = o_{k-3}(r^{-3/2})
	\;.\label{FOM::CanF}
\end{multline}
We have thus shown:

\begin{theorem}\label{RepThm}
Let $(M,g)$ be a three-dimensional smooth simply connected
manifold with a smooth connected compact boundary $\partial M$ and assume that $(M,g)$ admits a Killing vector field with periodic orbits. Furthermore, assume that $(M,g)$ has one asymptotically flat end where it satisfies \eqref{FO-1} for some $k \geq 5$. Then there exists a unique $m_1 > 0$ such that $M$ is diffeomorphic to $\RR^3\setminus I$ for some line interval $I$ of length $2m_1$, and, in cylindrical coordinates $(\rho,z,\varphi)$ of $\RR^3$ aligning so that $I = [-m_1,m_1]$ lies on the $z$-axis, the metric $g$ takes the form \eqref{Metric::CanF}, $\partial_\varphi$ is the rotational Killing vector field of $M$, and $U$, $\alpha$, $A_\rho$ and $B_z$ satisfy \eqref{CF::UDefV}, \eqref{CF::AlDefV}, \eqref{CF::AxisReg}, \eqref{CF::BrDefV}, \eqref{CF::AzDefV} and \eqref{FOM::CanF}.
\end{theorem}

\begin{remark}
The above analysis can be carried out with some additional work
to take care of the case where $\partial M$ is disconnected.
The only delicate point is the construction of the coordinates
$(\rho_S,z_S)$ such that, in the $(\rho_S,z_S)$-plane,
$\partial M$ corresponds to a union of a finite number of
disjoint circles. An alternative way is to first construct the
$(\rho,z)$ coordinates as in \cite[Section 6.3]{ChCo}, and use
our analysis here to derive the behaviour near each component
of $\partial M$ of the functions of interest. This approach
simplifies the analysis in \cite[Section 6.5]{ChCo}.
\end{remark}

%++++++++++++++++++++++++++++++++++++++++%
\subsection{The constant $m_1$}

We showed earlier that $m_1$ is uniquely determined by the geometry of $(M,g)$. Here we give a more explicit description of $m_1$.

Recall that $Q^{(2)}$ is represented by $\Omsh$ in
$(\rsh,\zsh)$-coordinates and that $m_1$ can be expressed in
terms of the Robin constant $\gamma(\partial\Omsh)$ of
$\partial\Omsh$ by \eqref{CF::m1Def}. By definition, if $\Gamma
= \Gamma_{\partial\Omsh}$ is the unique harmonic function in
$\Csh$ (with the flat metric) which vanishes at $\partial\Omsh$
and is asymptotic to $\frac{1}{2}\log(\rsh^2 + \zsh^2)$ at
infinity, then
\[
\Gamma(\rsh,\zsh) = \frac{1}{2}\log(\rsh^2 + \zsh^2) + \gamma(\partial\Omsh) + O((\rsh^2 + \zsh^2)^{-1/2}).
\]

Let $\{y^1, y^2, y^3 = \varphi\}$ be a coordinate system on $M$
such that $\{y^1,y^2\}$ is a coordinate system on $Q^{(2)}$. In
the sequel, indices $a$ and $b$ range over $\{1,2\}$, while Greek
indices range over $\{1,2,3\}$. The induced quotient metric on
$Q^{(2)}$ is given by
\[
q_{ab} = g_{ab} - {}^K g^{\varphi\varphi}\,g_{\varphi\,a}\,g_{\varphi\,b}
	\;,
\]
where ${}^K g^{\varphi\varphi} = \frac{1}{g_{\varphi\varphi}}$. Note that, by \eqref{MetricSh::CanF},
\[
q = e^{-2\Ush + 2\ash}(d\rsh^2 + d\zsh^2)
	\;.
\]
Thus, as a function on $Q^{(2)}$, $\Gamma$ is harmonic with respect to the metric $q$, i.e.
\[
\partial_{y^a} \Big(\sqrt{\det q}\,q^{ab}\,\partial_{y^b} \Gamma\Big) = 0 \text{ in } \mathring{Q}^{(2)}
	\;.
\]
Since $\Gamma$ is $\varphi$-independent and $\partial_\varphi$ is Killing, this implies that as a function on $M$, $\Gamma$ satisfies
\begin{equation}
\partial_{y^\mu} \left(\sqrt{\frac{\det g}{g_{\varphi\varphi}}}\,g^{\mu\nu}\,\partial_{y^\nu} \Gamma\right) = 0 \text{ in } \ringM\setminus \mcA
	\;.\label{GammaPrepEqn}
\end{equation}
We thus conclude that $\Gamma$ satisfies
\begin{equation}
\left\{\begin{array}{ll}
L\,\Gamma := \Delta_g \Gamma - \frac{1}{2}\,g\big(\nabla_g \log g_{\varphi\varphi},\nabla_g \Gamma\big) = 0
	& \text{ in } \ringM\setminus \mcA
	\;,\\
\Gamma = 0
	& \text{ on } \partial M
	\;,\\
\Gamma = \log r + O(1)
	&\text{ as } r \rightarrow \infty
	\;,
\end{array}\right.\label{GammaEqn}
\end{equation}
where $r$ is the coordinate radius in the asymptotic region. Moreover, by construction, $\Gamma$ is the unique solution to
\eqref{GammaEqn}  satisfying $\partial_\varphi \Gamma \equiv
0$.

We thus have:

\begin{proposition}\label{m1Value?}
The constant $m_1$ is given by
\begin{equation}
m_1 = 2\,\exp\Big(-\lim_{r\rightarrow\infty} (\Gamma - \log r)\Big)
	\;,\label{CF::m1DefV}
\end{equation}
where $\Gamma$ is the unique axially symmetric smooth solution
to \eqref{GammaEqn}.
\end{proposition}

%----------------------------------------------------------------------------%
\section{The ADM mass}\label{Sec::ADM}

In this section, we compute the ADM mass $m$ of $g$ as a volume integral over $\RR^3 \setminus B(0,\frac{m_1}{2})$ and then use it to prove Theorem \ref{MainThm}. We have
\[
m = \lim_{R \rightarrow \infty} \frac{1}{16\pi}\int_{S_R} (g_{ij,j} - g_{jj,i})\nu_i\,d\sigma,
\]
where the metric components are computed in a coordinate system
satisfying \eqref{FO-1}-\eqref{FO-3},
$d\sigma$ is the surface area form on $S_R$, and $S_R$ can be
taken to be any piecewise differentiable surface homologous to
a coordinate sphere of radius $R$ with
\[
\inf\{ r(p): p \in S_R\} \rightarrow _{R\to \infty}\infty.
\]
That the ADM mass is well-defined is well-known, see
\cite{ChErice,Bartnik}.

%+++++++++++++++++++++++++++++++++++%

\subsection{Mass in pseudo-spherical coordinates}

Define
\[
x^1 = x_S = \rho_S\,\cos\varphi
	\;, \qquad x^2 = y_S = \rho_S\,\sin\varphi
	\;, \qquad x^3 = z_S
	\;.
\]
Using \eqref{FOMS::CanF}, we can write the metric \eqref{MetricS::CanF} as
\begin{align}
    g
	= \;& e^{-2U_S}(dx_S^2 + dy_S^2) + \frac{e^{-2U_S}(e^{2\alpha_S} -1)}{\rho_S^2}(x_S\,dx_S + y_S\,dy_S)^2
 \nonumber
\\
    &
     + e^{-2U_S+ 2\alpha_S}\,dz_S^2+ 2(x_S\,dy_S - y_S\,dx_S)\,(\bar B_S\,(x_S\,dx_S + y_S\,dy_S)
    \nonumber
\\
		&
		+ \bar A_S\,dz_S) + o_1(r^{-1})
	\;.\label{MSCF::AsForm}
\end{align}
Here $r$ denotes the coordinate radius, $r = \sqrt{x_S^2 + y_S^2 + z_S^2}$.

In the following computation, $S_R$ is the sphere of coordinate radius $r := \sqrt{x_S^2 + y_S^2 + z_S^2} = R$. Obviously, the error terms in \eqref{MSCF::AsForm} has no contribution to the mass integral. A straightforward computation using \eqref{MetricS::CanF} shows that the terms involving $\bar B_S$ and $\bar A_S$ give also zero contribution to the mass integral.

The rest of the mass integrand is then found to be
\begin{align*}
&\Big\{
	\partial_{y_S}\Big(e^{-2U_S}(e^{2\alpha_S} - 1)\,\frac{x_S\,y_S}{\rho_S^2}\Big)\\
	&\qquad\qquad- \partial_{x_S}\Big(e^{-2U_S} + e^{-2U_S}(e^{2\alpha_S} - 1)\,\frac{y_S^2}{\rho_S^2}\Big)
		- \partial_{x_S}\,e^{-2U_S + 2\alpha_S}
	\Big\}\frac{x_S}{r}\\
&+\Big\{
	\partial_{x_S}\Big(e^{-2U_S}(e^{2\alpha_S} - 1)\,\frac{x_S\,y_S}{\rho_S^2}\Big)\\
	&\qquad\qquad- \partial_{y_S}\Big(e^{-2U_S} + e^{-2U_S}(e^{2\alpha_S} - 1)\,\frac{x_S^2}{\rho_S^2}\Big)
		- \partial_{y_S}\,e^{-2U_S + 2\alpha_S}
	\Big\}\frac{y_S}{r}\\
&+\Big\{
	- \partial_{z_S}\Big(e^{-2U_S} + e^{-2U_S}(e^{2\alpha_S} - 1)\,\frac{x_S^2}{\rho_S^2}\Big)\\
	&\qquad\qquad  - \partial_{z_S}\Big(e^{-2U_S} + e^{-2U_S}(e^{2\alpha_S} - 1)\,\frac{y_S^2}{\rho_S^2}\Big)
	\Big\}\frac{z_S}{r}
	\;.
\end{align*}
Upon simplifying this gives
\begin{align*}
&\Bigg\{
	\partial_{y_S}\Big(e^{-2U_S}(e^{2\alpha_S} - 1)\,\frac{x_S\,y_S}{\rho_S^2}\Big)\\
	&\qquad\qquad - \partial_{x_S}\Big(e^{-2U_S}\,\frac{x_S^2}{\rho_S^2} + e^{-2U_S + 2\alpha_S}\Big(1 + \frac{y_S^2}{\rho_S^2}\Big)\Big)
		\Bigg\}\frac{x_S}{r}\\
&+\Bigg\{
	\partial_{x_S}\Big(e^{-2U_S}(e^{2\alpha_S} - 1)\,\frac{x_S\,y_S}{\rho_S^2}\Big)\\
	&\qquad\qquad- \partial_{y_S}\Big(e^{-2U_S}\,\frac{y_S^2}{\rho_S^2} + e^{-2U_S + 2\alpha_S}\Big(1 + \frac{x_S^2}{\rho_S^2}\Big)
	\Bigg\}\frac{y_S}{r}\\
& - \partial_{z_S}\Big(e^{-2U_S}(e^{2\alpha_S} + 1)\Big)\frac{z_S}{r}
	\;.
\end{align*}
Expanding using \eqref{FOMS::CanF} we obtain
\[
\partial_R(2U_S - \alpha_S)
	+ \frac{2}{r}\,\alpha_S
	+ o(r^{-2})
	\;.
\]

We thus arrive at
\begin{equation}
m = \frac{1}{4\pi} \lim_{R\rightarrow \infty} \Big\{\int_{S_R} \partial_r\Big(U_S - \frac{1}{2}\alpha_S\Big) \,d\sigma + \frac{1}{2R}\int_{S_R}\alpha_S\,d\sigma \Big\}
	\;.\label{ADMMass-RoundSphere}
\end{equation}
(This is similar to a formula derived in~\cite{ChUone}, but the integrations are over different sets, which requires the new derivation above. The current expression is more convenient for our purposes.)

To proceed, we recall a formula for the scalar curvature on $M$
from \cite{GibbonsHolzegel},
\begin{multline}
R_g = 4\,e^{2U_S - 2\alpha_S}\Big[\Delta\Big(U_S - \frac{1}{2}\alpha_S\Big) - \frac{1}{2}|\nabla U_S|^2 + \frac{1}{2\rho_S}\partial_{\rho_S} \alpha_S\\
		- \frac{1}{8}\rho_S^2\,e^{-2\alpha_S}(\rho_S\,\partial_{z_S} \bar B_S - \partial_{\rho_S} \bar A_S)^2\Big].
\label{ScalarCurv-rSzS}
\end{multline}
Here $\Delta$ and $\nabla$ are the Laplacian and the gradient
operator taken with respect to the flat metric in $\RR^3$.

Using \eqref{ScalarCurv-rSzS}, we can convert \eqref{ADMMass-RoundSphere} into volume integral form. Note that if $\Phi$ is a function defined on $\RR^3 \setminus B(0,\frac{m_1}{2})$ satisfying
\begin{equation}
\Phi > 0 \text{ in } \RR^3 \setminus B(0,\frac{m_1}{2})
	\;, \text{ and } \Phi = 1 + o(r^{-1/2}) \text{ as } r \rightarrow \infty
	\;,\label{PhiDecayW}
\end{equation}
then
\begin{equation}
\lim_{R \rightarrow \infty} \int_{S_R} \partial_r \Big(U_S - \frac{1}{2}\alpha_S\Big)\,d\sigma =
	\lim_{R \rightarrow \infty} \int_{S_R} \Phi\,\partial_r \Big(U_S - \frac{1}{2}\alpha_S\Big)\,d\sigma
	\;.\label{GInf}
\end{equation}
By the divergence theorem, we have
\begin{align*}
&\int_{S_R} \Phi\,\partial_r\Big(U_S - \frac{1}{2}\alpha_S\Big)\,d\sigma\\
	&\qquad= \int_{B(0,R) \setminus B(0,\frac{m_1}{2})} \Big[\nabla \Phi \cdot \nabla \Big(U_S - \frac{1}{2}\alpha_S\Big) + \Phi\,\Delta\Big(U_S - \frac{1}{2}\alpha_S\Big)\Big]\,d^3x\\
		&\qquad\qquad +\int_{\partial B(0,\frac{m_1}{2})} \Phi\,\partial_r\Big(U_S - \frac{1}{2}\alpha_S\Big)\,d\sigma
	\;.
\end{align*}
Hence, by \eqref{ScalarCurv-rSzS},
\begin{align}
\int_{S_R} \Phi\,\partial_r\Big(U_S - \frac{1}{2}\alpha_S\Big)\,d\sigma
	=&
     \int_{B(0,R) \setminus B(0,\frac{m_1}{2})}\Big\{\nabla \Phi \cdot \nabla \Big(U_S - \frac{1}{2}\alpha_S\Big)\nonumber\\
			&\qquad + \frac{1}{2}\,\Phi\,|\nabla U_S|^2 - \frac{1}{2\rho_S}\partial_{\rho_S} \alpha_S\,\Phi\nonumber\\
			&\qquad  + \frac{1}{4}\,e^{-2U_S + 2\alpha_S}\,\Phi\,R_g \nonumber\\
			&\qquad  + \frac{1}{8}\rho_S^2\,e^{-2\alpha_S}\,\Phi(\rho_S\,\partial_{z_S} \bar B_S - \partial_{\rho_S} \bar A_S)^2\Big\}\,d^3x\nonumber\nonumber\\
		&+\int_{\partial B(0,\frac{m_1}{2})} \Phi\,\partial_r\Big(U_S - \frac{1}{2}\alpha_S\Big)\,d\sigma
	\;.\label{FirstGuy}
\end{align}
To get rid of the terms involving gradients of $\alpha_S$ we choose $\Phi$ to
satisfy
\begin{equation}
\frac{1}{\rho_S}\divop(\rho_S\,\nabla\Phi) = \divop\big(\nabla\Phi + \Phi\,\nabla \log\rho_S\big) = 0 \text{ in } \RR^3 \setminus B(0,\frac{m_1}{2})
	\;.\label{PhiChoice}
\end{equation}
Note that if we view $\Phi$ as a function defined in $\RR^4 \setminus B(0,\frac{m_1}{2})$ invariant under $SO(2)$ and assume that $\Phi$ is locally bounded, then
\begin{equation}
\Delta^{(4)}\Phi = 0 \text{ in } \RR^4 \setminus B(0,\frac{m_1}{2})
	\;,\label{PhiChoice'}
\end{equation}
In particular, this implies that $\frac{1}{\rho_S}\partial_{\rho_S}\Phi$ is locally bounded, and $\partial\Phi=O(r^{-3})$ for large $r$. Thus, as $\alpha_S$ vanishes wherever $\rho_S = 0$, an application of the divergence theorem gives
\begin{align*}
&\int_{B(0,R) \setminus B(0,\frac{m_1}{2})} \Big[\nabla \Phi\cdot \nabla \alpha_S + \frac{1}{\rho_S}\,\partial_{\rho_S}\alpha_S\,\Phi\Big]\,d^3x\\
	&\qquad\qquad = \int_{B(0,R) \setminus B(0,\frac{m_1}{2})} \nabla\alpha_S \cdot \big(\nabla \Phi + \Phi\,\nabla\log \rho_S\big)\,d^3x\\
	&\qquad\qquad = \int_{S_R} \alpha_S\,(\partial_r \log \rho_S+O(R^{-3}))\,d\sigma\\
		&\qquad\qquad\qquad\qquad - \int_{\partial B(0,\frac{m_1}{2})} \alpha_S\,\big(\partial_r \Phi + \Phi\,\partial_r \log \rho_S\big)\,d\sigma\\
	&\qquad\qquad = \frac{1}{R}\int_{S_R} \alpha_S\,d\sigma
 +o(R^{-3/2})	- \int_{\partial B(0,\frac{m_1}{2})} \alpha_S\,\big(\partial_r \Phi + \frac{2}{m_1}\Phi\big)\,d\sigma
	\;.
\end{align*}
Substituting the above into \eqref{FirstGuy} yields
\begin{align*}
&\int_{S_R} \Phi\,\partial_r\Big(U_S - \frac{1}{2}\alpha_S\Big)\,d\sigma\\
	&\qquad= \int_{B(0,R) \setminus B(0,\frac{m_1}{2})}\Big\{\nabla \Phi \cdot \nabla U_S  + \frac{1}{2}\,\Phi\,|\nabla U_S|^2\\
			&\qquad\qquad\qquad + \frac{1}{4}\,e^{-2U_S +
    2\alpha_S}\,\Phi\,R_g
\\
			&\qquad\qquad\qquad + \frac{1}{8}\rho_S^2\,e^{-2\alpha_S}\,\Phi(\rho_S\,\partial_{z_S} \bar B_S - \partial_{\rho_S} \bar A_S)^2\Big\}\,d^3x\\
		&\qquad\qquad - \frac{1}{2R}\int_{S_R} \alpha_S\,d\sigma
 +o(R^{-3/2})	
\\
		&\qquad\qquad +\int_{\partial B(0,\frac{m_1}{2})}
  \Big\{\Phi\,\partial_r\Big(U_S - \frac{1}{2}\alpha_S\Big) + \frac{1}{2}\alpha_S\,\big(\partial_r \Phi + \frac{2}{m_1}\Phi\big)\Big\}\,d\sigma
	\;.
\end{align*}
Recalling \eqref{ADMMass-RoundSphere} and \eqref{GInf}, we arrive at
\begin{align}
m
	&= \frac{1}{4\pi}\int_{\R^3 \setminus B(0,\frac{m_1}{2})}\Big\{\nabla \Phi \cdot \nabla U_S  + \frac{1}{2}\,\Phi\,|\nabla U_S|^2\nonumber\\
			&\qquad\qquad + \frac{1}{4}\,e^{-2U_S + 2\alpha_S}\,\Phi\,R_g \nonumber\\
			&\qquad\qquad + \frac{1}{8}\rho_S^2\,e^{-2\alpha_S}\,\Phi(\rho_S\,\partial_{z_S} \bar B_S - \partial_{\rho_S} \bar A_S)^2\Big\}\,d^3x\nonumber\\
		&\qquad + \frac{1}{4\pi} \int_{\partial B(0,\frac{m_1}{2})} \Big\{\Phi\,\partial_r\Big(U_S - \frac{1}{2}\alpha_S\Big) + \frac{1}{2}\alpha_S\,\big(\partial_r \Phi + \frac{2}{m_1}\Phi\big)\Big\}\,d\sigma
	\;.\label{ADMVol-RoundSphere-1}
\end{align}

Next, if $\Psi$ is a function defined on $\RR^3 \setminus B(0,\frac{m_1}{3})$ such that
\begin{equation}
\left\{\begin{array}{ll}
\Delta\Psi = 0 & \text{ in } \RR^3 \setminus \bar B(0,\frac{m_1}{3})	\;,\\
\Psi = {\rm Const} + O(r^{-1})  &\text{ as } r\rightarrow \infty	\;,
\end{array}\right.
	\label{PsiChoice}
\end{equation}
then
\begin{equation}
\int_{\R^3\setminus B(0,\frac{m_1}{2})} \nabla \Psi \cdot \nabla U_S\,d^3x
	= -\int_{\partial B(0,\frac{m_1}{2})} U_S\,\partial_r\Psi \,d\sigma
	\;.\label{PsiProp}
\end{equation}

Using the above identity in \eqref{ADMVol-RoundSphere-1} yields
\begin{align}
m
	=& \frac{1}{4\pi}\int_{\R^3 \setminus B(0,\frac{m_1}{2})}\Big\{\frac{1}{2}\,\Phi\,|\nabla U_S|^2 + \nabla U_S \cdot \nabla(\Phi - \Psi)\nonumber\\
			&\qquad\qquad + \frac{1}{4}\,e^{-2U_S + 2\alpha_S}\,\Phi\,R_g \nonumber\\
			&\qquad\qquad + \frac{1}{8}\rho_S^2\,e^{-2\alpha_S}\,\Phi(\rho_S\,\partial_{z_S} \bar B_S - \partial_{\rho_S} \bar A_S)^2\Big\}\,d^3x\nonumber\\
		&+ \frac{1}{4\pi} \int_{\partial B(0,\frac{m_1}{2})} \Big\{\Phi\,\partial_r\Big(U_S - \frac{1}{2}\alpha_S\Big) + \frac{1}{2}\alpha_S\,\big(\partial_r \Phi
			+ \frac{2}{m_1}\Phi\big)\nonumber\\
			&\qquad\qquad - U_S\,\partial_r\Psi\Big\}\,d\sigma
	\;.\label{ADMVol-RoundSphere-2}
\end{align}

To conclude, we have shown:
\begin{Proposition}
Under the hypotheses of Theorem \ref{RSRepThm} and
\eqref{FO-2}, the ADM mass of $(M,g)$ is well-defined and
satisfies \eqref{ADMVol-RoundSphere-2} for any $\Phi$ and $\Psi$ satisfying \eqref{PhiDecayW},
\eqref{PhiChoice} and \eqref{PsiChoice}.
\end{Proposition}

We shall show below how appropriate choices of $\Phi$ and $\Psi$ allow one to control the mass.

For further reference we note:

\begin{corollary}\label{C28V10.1}
If $(\Phi_1,\Psi_1)$ and $(\Phi_2,\Psi_2)$ satisfy \eqref{PhiDecayW}, \eqref{PhiChoice} and \eqref{PsiChoice} then
\begin{align}
0
	&= \frac{1}{4\pi}\int_{\R^3 \setminus B(0,\frac{m_1}{2})}\Big\{\frac{1}{2}\,(\Phi_1 - \Phi_2)\,|\nabla U_S|^2 + \nabla U_S \cdot \nabla(\Phi_1 - \Phi_2 - \Psi_1 + \Psi_2)\nonumber\\
			&\qquad\qquad + \frac{1}{4}\,e^{-2U_S + 2\alpha_S}\,(\Phi_1 - \Phi_2)\,R_g \nonumber\\
			&\qquad\qquad + \frac{1}{8}\rho_S^2\,e^{-2\alpha_S}\,(\Phi_1 - \Phi_2)\,(\rho_S\,\partial_{z_S} \bar B_S - \partial_{\rho_S} \bar A_S)^2\Big\}\,d^3x\nonumber\\
		&\qquad + \frac{1}{4\pi} \int_{\partial B(0,\frac{m_1}{2})} \Big\{(\Phi_1 - \Phi_2)\,\partial_r\Big(U_S - \frac{1}{2}\alpha_S\Big)\nonumber\\
        &\qquad\qquad + \frac{1}{2}\alpha_S\,\big(\partial_r \Phi_1 + \frac{2}{m_1}\Phi_1 - \partial_r \Phi_2 - \frac{2}{m_1}\Phi_2\big)\nonumber\\
        &\qquad\qquad - U_S\,(\partial_r\Psi_1 - \partial_r \Psi_2)\Big\}\,d\sigma
	\;.\label{ADMVol-RoundSphere-2x}
\end{align}
\end{corollary}

%+++++++++++++++++++++++++++++++++++%
\subsection{Lower bound for the ADM mass}

In this section, we prove Theorem \ref{MainThm}.
We now assume that
\begin{equation}
R_g \geq 0 \text{ in } M
	\;,\label{NonnegSCurv}
\end{equation}
together with a Riemannian version of the condition that $\partial M$ is weakly outer trapped, namely:
\begin{equation}
\text{the mean curvature of $\partial M$ is non-positive}
	\;.\label{WOuterTrapped}
\end{equation}
Here the mean curvature is computed with respect to the normal pointing towards $M$. By a direct computation, \eqref{WOuterTrapped} is equivalent to
\begin{equation}
\partial_r \big(U_S - \frac{1}{2}\alpha_S\big) \geq \frac{2}{m_1} \text{ on } \partial B(0,\frac{m_1}{2})
	\;.\label{WOuterTrapped-rSzS}
\end{equation}

\medskip
\noindent{\sc Proof of Theorem \ref{MainThm}:} Under \eqref{NonnegSCurv} and \eqref{WOuterTrapped},
\eqref{ADMVol-RoundSphere-2} implies, keeping in mind that $\Phi$ is positive, and
completing the square in the volume integral when passing from
the first to the second inequality,
\begin{align}
m
	&\geq \frac{1}{4\pi}\int_{\R^3 \setminus B(0,\frac{m_1}{2})}\Big\{\frac{1}{2}\,\Phi\,|\nabla U_S|^2 + \nabla U_S \cdot \nabla(\Phi - \Psi)\Big\}\,d^3x\nonumber\\
		&\qquad + \frac{1}{4\pi} \int_{\partial B(0,\frac{m_1}{2})} \Big\{\Phi\,\frac{2}{m_1} + \frac{1}{2}\alpha_S\,\big(\partial_r \Phi + \frac{2}{m_1}\Phi\big) - U_S\,\partial_r \Psi\Big\}\,d\sigma\nonumber\\
	&\geq -\frac{1}{8\pi}\int_{\R^3 \setminus B(0,\frac{m_1}{2})} \frac{1}{\Phi}\,|\nabla (\Phi - \Psi)|^2\,d^3x\nonumber\\
		&\qquad + \frac{1}{4\pi} \int_{\partial B(0,\frac{m_1}{2})} \Big\{\Phi\,\frac{2}{m_1} + \frac{1}{2}\alpha_S\,\big(\partial_r \Phi + \frac{2}{m_1}\Phi\big) - U_S\,\partial_r \Psi\Big\}\,d\sigma
	\;.\label{ADMRS::Bnd1}
\end{align}

To continue, we specialize the choice of $\Phi$ and $\Psi$ by taking
\[
\Psi \equiv 1 \text{ and }\Phi \equiv 1 + \frac{m_1^2}{4r^2}
	\;.
\]
Then \eqref{ADMRS::Bnd1} gives
\begin{align}
m
	&\geq m_1 - \frac{1}{8\pi}\int_{\RR^3 \setminus B(0,\frac{m_1}{2})} \frac{m_1^4}{r^4(4r^2 + m_1^2)}\,d^3x\nonumber\\
	&= \frac{\pi}{4}m_1
	\;.\label{ADMRS::Bnd4}
\end{align}

Next, assume that $m = \frac{\pi}{4}\,m_1$. Then, we must have
\begin{align}
R_g
	&\equiv \rho_S\,\partial_{z_S} \bar B_S - \partial_{\rho_S} \bar A_S \equiv 0 \text{ in } \RR^3 \setminus B(0,\frac{m_1}{2})
	\;,\label{Rig1}\\
\nabla U_S
	&\equiv -\frac{1}{\Phi }\nabla(\Phi -\Psi)= -\frac 1 {\Phi} \nabla \Phi \text{ in } \RR^3 \setminus B(0,\frac{m_1}{2})
	\;,\label{Rig2}\\
\partial_r\Big(U_S - \frac{1}{2}\alpha_S\Big)
	&\equiv \frac{2}{m_1} \text{ on } \partial B(0,\frac{m_1}{2})
	\;.\label{Rig3}
\end{align}
By \eqref{FOMS::CanF}, the second relation implies that
\[
 \nabla U_S \equiv \frac{2m_1^2}{r(4r^2 + m_1^2)}\partial_r \ \text{ in }\  \RR^3 \setminus B(0,\frac{m_1}{2})
	\;,
\]
and so, since $U_S$ is assumed to asymptote to zero at infinity,
\begin{equation}
U_S \equiv \log\frac{4r^2}{4r^2 + m_1^2} \ \text{ in } \ \RR^3 \setminus B(0,\frac{m_1}{2})
	\;.
    \label{RigUS}
\end{equation}

Taking \eqref{ScalarCurv-rSzS}, \eqref{Rig3} and \eqref{RigUS} into account we get
\[
\left\{\begin{array}{ll}
\Delta \alpha_S - \frac{1}{\rho_S}\,\partial_{\rho_S}\alpha_S = -\frac{16m_1^2 }{(4r^2+m_1^2)^2} < 0
	& \text{ in } \RR^3 \setminus \bar B(0,\frac{m_1}{2})
	\;,\\
\partial_r \alpha_S = 0
	& \text{ on } \partial B(0,\frac{m_1}{2})
	\;,\\
\alpha_S = o(r^{-1/2})
	& \text{ as } r \rightarrow \infty
	\;.
\end{array}\right.
\]
Since $\alpha_S$ is $\varphi$-independent, this implies
\[
\left\{\begin{array}{ll}
\partial_{\rho_S}^2 \alpha_S + \partial_{z_S}^2 \alpha_S < 0
	& \text{ in } \RR^2 \setminus \bar D(0,\frac{m_1}{2})
	\;,\\
\partial_r \alpha_S = 0
	& \text{ on } \partial D(0,\frac{m_1}{2})
	\;,\\
\alpha_S = o(r^{-1/2})
	& \text{ as } r \rightarrow \infty
	\;.
\end{array}\right.
\]
This is impossible by Hadamard's Three-Circle Theorem, proving that the equality cannot hold in \eqref{ADMRS::Bnd4}. We conclude the proof of Theorem~\ref{MainThm}.
\eproof

\appendix

\section{A remark on the axisymmetric Penrose inequality}
 \label{A10V10.1}

In~\cite{GibbonsHolzegel} a proof of the Penrose inequality for
axisymmetric initial data sets with positive scalar curvature
has been given, under however undesirably stringent conditions
on the geometry near the horizon. It seems therefore of
interest to attempt to remove the overly restrictive
conditions. In particular one can enquire whether our arguments
above can be adapted to obtain the Penrose inequality. In this
appendix we provide an argument that gives a result stronger
than that in~\cite{GibbonsHolzegel}, but fails to provide the
full Penrose inequality.

We will always assume \eqref{NonnegSCurv}, i.e $R_g \geq 0$ in $M$. Furthermore, we will assume that
\begin{equation}
\text{$\partial M$ is minimal, i.e. $\partial_r \Big(U_S - \frac{1}{2}\alpha_S\Big) = \frac{2}{m_1}$ on $\partial B(0,\frac{m_1}{2})$.}
\label{App::MinBdry}
\end{equation}

By the first inequality in \eqref{ADMRS::Bnd1}, we have
\begin{align}
m
	&\geq \frac{1}{4\pi}\int_{\RR^3 \setminus B(0,\frac{m_1}{2})}\Big\{\frac{1}{2}\,\Phi\,|\nabla U_S|^2 + \nabla U_S \cdot \nabla(\Phi - \Psi)\Big\}\,d^3x\nonumber\\
		&\qquad + \frac{1}{4\pi} \int_{\partial B(0,\frac{m_1}{2})} \Big\{\frac{2}{m_1}\,\Phi + \frac{1}{2}\alpha_S\,\big(\partial_r \Phi + \frac{2}{m_1}\Phi\big) - U_S\,\partial_r\Psi\Big\}\,d\sigma
	\label{App::mLB}
\end{align}
for any $\Phi$ and $\Psi$ satisfying \eqref{PhiDecayW}, \eqref{PhiChoice} and \eqref{PsiChoice}. Moreover, this is an equality iff $R_g \equiv 0 \equiv \rho_S\,\partial_{z_S} \bar B_S - \partial_{\rho_S} \bar A_S$.

Let $A$ be the area of $\partial M$. Then
\begin{equation}
A = \int_{\partial B(0,\frac{m_1}{2})} e^{\alpha_S - 2U_S}\,d\sigma
	\;.\label{App::Area}
\end{equation}

According to Bray, Huisken, and Ilmanen~\cite{HI2,Bray:preparation2} one
has
\bel{Pineq}
 m \ge \sqrt{\frac A {16 \pi}}
 \;.
\ee

Hence, under the stated hypotheses and that $\R^3\setminus B(0,\frac{m_1}{2})$ with the metric
\eqref{MetricS::CanF} contains no compact minimal surfaces other than its boundary, one would
naively expect that it must hold that
\begin{align}
&J_{\Phi,\Psi}(U_S,\alpha_S)\nonumber\\
	&\qquad:= \frac{1}{4\pi}\int_{\RR^3 \setminus B(0,\frac{m_1}{2})}\Big\{\frac{1}{2}\,\Phi\,\big|\nabla U_S + \frac{1}{\Phi}\nabla(\Phi - \Psi)|^2 - \frac{1}{2\Phi}|\nabla(\Phi - \Psi)|^2\Big\}\,d^3x\nonumber\\
		&\qquad\qquad + \frac{1}{4\pi} \int_{\partial B(0,\frac{m_1}{2})} \Big\{\frac{2}{m_1}\,\Phi + \frac{1}{2}\alpha_S\,\big(\partial_r \Phi + \frac{2}{m_1}\Phi\big) - U_S\,\partial_r\Psi\Big\}\,d\sigma\nonumber\\
		&\qquad\qquad - \sqrt{\frac{1}{16\pi}\int_{\partial B(0,\frac{m_1}{2})} e^{\alpha_S - 2U_S}\,d\sigma}
			\geq 0
	\;,\label{App::Hope}
\end{align}
for some well-chosen $\Phi$ and $\Psi$. Moreover equality should only hold for the Schwarzschild solution.

For a fixed $m_1$ this is thus a variational inequality: if the
infimum over $U_S$ and $\alpha_S$ as described above of
$J_{\Phi,\Psi}(U_S,\alpha_S)$ is zero, then the axisymmetric
Riemannian Penrose inequality would follow.

A natural choice for $\Phi$ and $\Psi$ is to use functions which make the first volume integrand in \eqref{App::Hope} vanish for the Schwarzschild
solution:
\begin{equation}
\nabla \USSchw + \frac{1}{\Phi}\,\nabla(\Phi - \Psi) \equiv 0
	\;,\label{Specltn}
\end{equation}
where $\USSchw$ is the ``$U_S$'' of the Schwarzschildian slice,
\[
\USSchw = -2 \log \frac{2r + m_1}{2r}
	\;.
\]
This leads to $\Phi = 1 + \frac{a\,m_1^2}{4r}$ and $\Psi = \frac{b\,m_1}{2r}$. (Here we have used the equations \eqref{PhiChoice'} and \eqref{PsiChoice}.) Entering this into \eqref{Specltn}, we obtain $a = -1$ and $b = -2$.

There is a special case where the expected inequality holds:

\begin{proposition}\label{PosResult}
For any $(U_S,\alpha_S)$ satisfying the relevant hypotheses and
\begin{equation}
U_S - \frac{1}{2}\alpha_S \equiv C_H \geq -2\log 2 \text{ on } \partial B(0,\frac{m_1}{2})	
	\;,\label{PosResultCond}
\end{equation}
there holds
\[
J_{\Phi_*,\Psi_*}(U_S,\alpha_S) \geq 0
	\;,
\]
where $\Phi_* = 1 - \frac{m_1^2}{4r^2}$ and $\Psi_* = -\frac{m_1}{r}$. Moreover, equality holds if and only if the metric \eqref{MetricS::CanF} is that of a Schwarzschildian
slice.
\end{proposition}

It should be noted that the existence of admissible data
verifying \eqref{PosResultCond} (other than the Schwarzschildian slice) is not clear. We also note that
the requirement that $\partial M$ be the outermost minimal surface is not necessary, but rather $\partial M$ being merely weakly outer trapped is sufficient.

Proposition~\ref{PosResult} should be compared with a result in~\cite{GibbonsHolzegel}, where equality in \eqref{PosResultCond} is assumed together with the supplementary requirement that $A$, as defined by \eqref{App::Area}, equals $16 \pi m_1^2$.

\medskip
\bproof We will only sketch the proof. Using the explicit form of $(\Phi_*,\Psi_*)$, one finds
\begin{align}
J_{\Phi_*,\Psi_*}(U_S,\alpha_S)
	&= -m_1(2\log2 - 1) - m_1\,C_H - \frac{1}{4}\,m_1\,e^{-C_H}\nonumber\\
		&\qquad\qquad + \frac{1}{8\pi} \int_{\RR^3 \setminus B(0,\frac{m_1}{2})} \Phi\,|V_S|^2\,d^3x
	\;,\label{JPP*}
\end{align}
where
\[
V_S = \nabla U_S + \frac{1}{\Phi_*} \nabla (\Phi_* - \Psi_*)
	\;.
\]

Next, set $\Xi = \frac{m_1^2}{2r^2}$. Applying Corollary \eqref{C28V10.1} to $(\Phi_1, \Psi_1) = \big(\Phi_* + \Xi,0)$ and $(\Phi_2,\Psi_2) = (\Phi_*,\Psi_*)$ and noting \eqref{NonnegSCurv} and \eqref{WOuterTrapped-rSzS}, we find
\begin{align*}
0
	&= \frac{1}{4\pi}\int_{\R^3 \setminus B(0,\frac{m_1}{2})}\Big\{\frac{1}{2}\,\Xi\,|\nabla U_S|^2 + \nabla U_S \cdot \nabla(\Xi + \Psi_*)\nonumber\\
			&\qquad\qquad + \frac{1}{4}\,e^{-2U_S + 2\alpha_S}\,\Xi\,R_g \nonumber\\
			&\qquad\qquad + \frac{1}{8}\rho_S^2\,e^{-2\alpha_S}\,\Xi\,(\rho_S\,\partial_{z_S} \bar B_S - \partial_{\rho_S} \bar A_S)^2\Big\}\,d^3x\nonumber\\
		&\qquad + \frac{1}{4\pi} \int_{\partial B(0,\frac{m_1}{2})} \Big\{\Xi\,\partial_r\Big(U_S - \frac{1}{2}\alpha_S\Big)\nonumber\\
        &\qquad\qquad + \frac{1}{2}\alpha_S\,\big(\partial_r \Xi + \frac{2}{m_1}\Xi \big)
             + U_S\,\partial_r \Psi \Big\}\,d\sigma\\
        	&\geq \frac{1}{4\pi}\int_{\R^3 \setminus B(0,\frac{m_1}{2})}\Big\{\frac{1}{2}\,\Xi\,|\nabla U_S|^2 + \nabla U_S \cdot \nabla(\Xi + \Psi_*)\Big\}\,d^3x\nonumber\\
		&\qquad + \frac{1}{4\pi} \int_{\partial B(0,\frac{m_1}{2})} \Big\{\frac{2}{m_1}\,\Xi
        + \frac{1}{2}\alpha_S\,\big(\partial_r \Xi + \frac{2}{m_1}\Xi \big)
             + U_S\,\partial_r \Psi_* \Big\}\,d\sigma
\end{align*}
Using the explicit expressions for $\Xi$ and $\Psi_*$, we then get
\begin{align*}
-  m_1 \,C_H
	&= \frac{1}{4\pi}\int_{\partial B(0,\frac{m_1}{2})} \Big[-\frac{1}{2}\alpha_S\,(\partial_r \Xi + \frac{2}{m_1}\Xi) - U_S\,\partial_r\Psi_*\Big]\,d\sigma\nonumber\\
	&\geq \frac{1}{4\pi}\int_{\partial B(0,\frac{m_1}{2})} \frac{2}{m_1}\Xi\,d\sigma\nonumber\\
	&\qquad\qquad + \frac{1}{4\pi}\int_{\RR^3 \setminus B(0,\frac{m_1}{2})} \Big\{ \nabla U_S \cdot \nabla (\Xi + \Psi_*)
			+ \frac{1}{2}|\nabla U_S|^2\,\Xi\Big\}\,d^3x
	\;.
\end{align*}
Recalling that $\nabla U_S = V_S - \frac{1}{\Phi_*}\,\nabla(\Phi_* - \Psi_*)$, we thus have
\begin{align*}
&-  m_1 \,C_H\\
	&\qquad \geq \frac{1}{4\pi}\int_{\partial B(0,\frac{m_1}{2})} \frac{2}{m_1}\Xi\,d\sigma\nonumber\\
        &\qquad + \frac{1}{4\pi}\int_{\RR^3 \setminus B(0,\frac{m_1}{2})} \Big\{ \frac{\Xi}{2\Phi_*^2}\,|\nabla(\Phi_* - \Psi_*)|^2 - \frac{1}{\Phi_*}\,\nabla(\Phi_* - \Psi_*) \cdot \nabla(\Xi + \Psi_*)\Big\}\,d^3x \nonumber\\
	   &\qquad + \frac{1}{4\pi}\int_{\RR^3 \setminus B(0,\frac{m_1}{2})} \Big\{ V_S \cdot \Big[\nabla (\Xi + \Psi_*) - \frac{\Xi}{\Phi_*}\,\nabla(\Phi_* - \Psi_*)\Big] + \frac{1}{2}\,|V_S|^2\,\Xi\Big\}\,d^3x
	\;.
\end{align*}
Using the explicit expressions for $\Phi_*$, $\Psi_*$ and $\Xi$ again, one arrives at
\begin{align}
-m_1\,C_H
	&\geq 2\,m_1\,\log 2 - \sqrt{ \frac{m_1}{8\pi} \int_{\RR^3 \setminus B(0,\frac{m_1}{2})} \Phi\,|V_S|^2\,d^3x}
	\;.\label{Bail*}
\end{align}

Define
\[
t = -C_H - 2\log 2, \text{ and } l := \frac{1}{8\pi\,m_1}\int_{\RR^3 \setminus B(0,\frac{m_1}{2})} \Phi\,|V_S|^2\,d^3x
	\;
\]
Then, by \eqref{JPP*} and \eqref{Bail*} and as $l \geq 0$ and $t \leq 0$ (by \eqref{PosResultCond}),
\begin{align*}
\frac{1}{m_1}\,J_{\Phi_*,\Psi_*}(U_S,\alpha_S)
	&= 1 + t - e^t + l\\
	&\geq 1 - \sqrt{l} - e^{-\sqrt{l}} + l
		\geq 0
	\;.
\end{align*}
This finishes the proof.
\eproof

The bad news for the above program arises from the following:

\begin{proposition}\label{NegResult}
There exists an ``admissible'' $(U_S,\alpha_S) = (U_S,0)$ such
that
\begin{enumerate}
\item $U_S - \frac{1}{2}\alpha_S \equiv C_H < -2\log 2$ on $\partial B(0,\frac{m_1}{2})$,
\item $J_{\Phi_*,\Psi_*}(U_S,\alpha_S) < 0$.
\end{enumerate}
\end{proposition}

\bproof The example is provided by conformally Schwarzschildian
metrics, i.e. $\alpha_S \equiv 0$. $U_S$ is given by
\[
U_S = U_S^{(k)} = -2 \log \Big[1 + \frac{6k+m_1^2}{2m_1r} -
\frac{k}{2r^2}\Big]
	\;,\qquad k \geq 0
	\;.
\]
For $k = 0$, this gives exactly the Schwarzschild metric. For $k > 0$, the scalar curvature is readily seen to be positive, as $e^{-U_S/2}$ is super-harmonic (with respect to the flat metric). One can check directly from \eqref{App::MinBdry} that $\partial M$ is minimal. In fact, for $k < \frac{m_1^2}{6}$, $\partial M$ is outermost minimal. (An easy way to see that is to check that, for those values of $k$, the coordinate spheres provide a foliations of $M$ by constant positive mean curvature surfaces.) The rest of the argument is to use \eqref{JPP*} to verify that $J_{\Phi_*,\Psi_*}(U_S,\alpha_S)$ is negative for sufficiently small $k > 0$.
\eproof

\bigskip
\noindent
{\sc Acknowledgments} We are grateful to Gustav Holzegel and Yanyan Li for many useful discussions. Part of this work was done while LN was a postdoc at the Oxford Centre for Nonlinear PDE. He wishes to thank the centre for its encouraging environment and its financial support through
the EPSRC Science and Innovation award to the centre (EP/E035027/1). He further wishes to thank the Vienna relativity group for hospitality and support during part of work on this paper.  PTC is  supported in part
by the Polish Ministry of Science and Higher Education grant Nr N N201 372736.

%----------------------------------------------------------------------------%

\def\polhk#1{\setbox0=\hbox{#1}{\ooalign{\hidewidth
  \lower1.5ex\hbox{`}\hidewidth\crcr\unhbox0}}}
  \def\polhk#1{\setbox0=\hbox{#1}{\ooalign{\hidewidth
  \lower1.5ex\hbox{`}\hidewidth\crcr\unhbox0}}} \def\cprime{$'$}
  \def\cprime{$'$} \def\cprime{$'$} \def\cprime{$'$}
\providecommand{\bysame}{\leavevmode\hbox to3em{\hrulefill}\thinspace}
\providecommand{\MR}{\relax\ifhmode\unskip\space\fi MR }
% \MRhref is called by the amsart/book/proc definition of \MR.
\providecommand{\MRhref}[2]{%
  \href{http://www.ams.org/mathscinet-getitem?mr=#1}{#2}
}
\providecommand{\href}[2]{#2}

%----------------------------------------------------------------------------%
\end{document}